\documentclass{article}

\usepackage{PRIMEarxiv}

\usepackage[utf8]{inputenc} % allow utf-8 input
\usepackage[T1]{fontenc}    % use 8-bit T1 fonts
\usepackage{hyperref}       % hyperlinks
\usepackage{url}            % simple URL typesetting
\usepackage{booktabs}       % professional-quality tables
\usepackage{amsfonts}       % blackboard math symbols
\usepackage{amsmath}        % advanced math formatting
\usepackage{nicefrac}       % compact symbols for 1/2, etc.
\usepackage{microtype}      % microtypography
\usepackage{lipsum}
\usepackage{fancyhdr}       % header
\usepackage{graphicx}       % graphics
\graphicspath{{media/}}     % organize your images and other figures under media/ folder
\usepackage{xcolor}
\usepackage[round]{natbib}

\usepackage[authormarkup=none]{changes}
\definechangesauthor[name={jl}, color=black]{jl}

\title{Subsurface Property Mapping using Google AlphaEarth Foundations
}

\author{
  Nori Nakata, Jingxiao Liu, Guodong Chen, Rie Nakata, Charuleka Varadharajan\\
  Lawrence Berkeley National Laboratory\\
  Berkeley, CA 94720\\
  \texttt{\{nnakata@lbl.gov\}} \\
}

\begin{document}
\maketitle

\begin{abstract}
Subsurface properties are essential for hazard assessment, energy and environmental management, and infrastructure resilience, but direct observations are sparse and uneven, motivating the use of surface observations as indirect constraints. Here we explore whether AlphaEarth embeddings can be applied to subsurface estimation despite indirect and non-unique physical links between surface and depth. We test this idea in two conterminous U.S. applications: shallow seismic site characterization using $V_S 30$ with embedding features alone and with conventional covariates (topographic slope and a tectonic-status indicator), and subsurface temperature reconstruction using embedding-based nonlinear regression. Across both applications, embedding-informed models recover spatially coherent, physically plausible patterns and outperform simpler baselines. The comparison also highlights a key difference: domain covariates materially stabilize $V_S 30$ regression, whereas temperature mapping relies primarily on embedding features. Overall, the results support the feasibility of foundation-model surface representations for regional surface-to-subsurface inference, while emphasizing the need for robust spatial validation under heterogeneous labels and uneven data coverage.
\end{abstract}

% keywords can be removed

\section{Introduction}
\par
Subsurface properties such as geology, rock types, seismic velocity, temperature, and fluid saturation are fundamental state variables for Earth and environmental sciences. They control how crustal materials deform and transmit waves, how fluids and heat move through the subsurface, and how natural and engineered systems evolve over time. Consequently, reliable subsurface maps are central to earthquake and infrastructure resilience, groundwater management, geothermal and hydrocarbon resource assessment, and long-term planning for critical energy and environmental systems.

\par
A core challenge is that direct subsurface measurements are sparse, expensive, and unevenly distributed, whereas surface observations are dense and repeated in space and time. In many settings, subsurface conditions are partly expressed at the surface through geomorphology, lithologic expression, vegetation patterns, moisture regime, drainage organization, and anthropogenic land modification. % \nori{citation}
% \nori{Need to be revised.} 
For example, low-relief sedimentary basins and coastal plains are often associated with lower near-surface stiffness, while rugged bedrock uplands are often associated with higher stiffness; similarly, regional thermal structure can co-vary with surface heat-flow signatures, tectonic setting, and hydrothermal expressions. These links are indirect and not one-to-one, but they motivate using surface information as probabilistic constraints on subsurface properties.

\par
This study tests that idea using AlphaEarth, a foundation-model embedding of multi-source Earth observation data.
\added[id=jl]{Although AlphaEarth has only recently been introduced, the literature already shows utility across a broad range of remote-sensing tasks. 
In the original AlphaEarth Foundations paper, Brown et al. demonstrated strong transfer across land-cover and land-use mapping, change detection, crop-type mapping, tree-genera and plantation classification, and biophysical regression targets such as evapotranspiration and surface emissivity \citep{BrownEtAl2025}. Subsequent published work has applied AlphaEarth embeddings to county-scale crop-yield estimation \citep{fang2026yield}, wetland vegetation mapping \citep{ryan2026wetland}, forest biomass prediction and GEDI-assisted biomass estimation \citep{jin2026biomass,pascual2026biomass}, and urban air-quality prediction \citep{alvarez2025air}. Collectively, these studies suggest that AlphaEarth can serve as a general-purpose geospatial representation for both classification and regression tasks, often reducing the need for extensive handcrafted preprocessing while retaining strong transferability across regions and in label-scarce settings.}
In this study, we treat AlphaEarth embeddings as compact descriptors of coupled surface conditions and evaluate whether they improve prediction of multiple subsurface targets when paired with task-specific learning models. Our goal is primarily methodological and research-oriented: to assess transferability, quantify performance gains relative to conventional baselines, and identify where embedding-based inference is robust versus uncertain. By comparing two applications (shallow seismic velocity and subsurface temperature), we aim to clarify both the promise and limitations of surface-to-subsurface inference for scalable regional mapping.

\section{AlphaEarth-Based Framework and Applications}
\subsection{AlphaEarth Foundations as a surface representation}

In this study, we use AlphaEarth Foundations as a compact representation of surface conditions for downstream subsurface mapping. 
AlphaEarth is a geospatial foundation model that produces a global ``embedding field,'' that is, a stack of 64-dimensional vectors defined at 10~m spatial resolution and learned from multi-source Earth observation data rather than from a small set of hand-crafted predictors \citep{BrownEtAl2025}. 
In the original work, the model is trained to reconcile information from optical imagery, radar, topography, LiDAR, climate reanalysis, gravity-derived hydrologic signals, and auxiliary land-cover and georeferenced text sources into a common latent space. 
A key property of the method is that the embedding represents a temporally summarized surface state over a chosen period, rather than a single image acquisition. 
The annual product released in Google Earth Engine is therefore well suited to applications such as those considered here, where the targets are expected to depend more strongly on persistent landscape structure than on short-lived surface fluctuations.
\par
AlphaEarth does not directly observe subsurface information, such as $V_{S}30$ or temperature at depth. 
Instead, it encodes coupled surface expressions of geology, geomorphology, vegetation, hydrology, land use, and climate, all of which can provide indirect constraints on subsurface structure and state. 
These relations are non-unique and scale dependent, so throughout this paper AlphaEarth is interpreted as a statistical description of landscape context rather than as a direct physical measurement of the subsurface. 
\par
The original AlphaEarth study also demonstrated strong transfer across sparse-label tasks, including land-cover and land-use mapping, change detection, crop mapping, and regression targets such as evapotranspiration and surface emissivity. 
That transfer setting closely matches the goal of this paper and motivates our use of AlphaEarth as an off-the-shelf representation. 
We do not retrain or fine-tune the foundation model itself; instead, we learn task-specific predictors on top of fixed embedding vectors. 
% At the same time, the role of AlphaEarth differs slightly between the two applications considered here. 
% For shallow seismic site characterization, physically interpretable predictors such as topographic slope and broad regional tectonic setting remain informative, so AlphaEarth is used as an augmenting representation alongside conventional covariates. 
% For subsurface temperature, where controls are broader and more nonlinear, the embedding serves as the primary geospatial descriptor. 
% In this way, AlphaEarth provides a common surface representation across both applications while still allowing the downstream models to remain task specific.
\par
\par
In this study, AlphaEarth Foundations is used as a fixed surface representation rather than as a retrained model. For a selected annual layer \(y\) from the Google Earth Engine collection \texttt{GOOGLE/SATELLITE\_EMBEDDING/V1/ANNUAL}, the embedding field assigns to each surface location a 64-dimensional vector at nominal 10~m resolution. Let \(i=1,\ldots,n\) index samples, where a sample may be either a $V_S$30 station or a temperature-grid centroid, and let
$\mathbf{x}_i=(\lambda_i,\phi_i)$
denote its longitude--latitude coordinate. For embedding channel \(k=1,\ldots,64\), we denote the local AlphaEarth vector by
\begin{equation}
\mathbf{e}(\mathbf{x}_i,y)
=
\left[
e_1(\mathbf{x}_i,y), e_2(\mathbf{x}_i,y), \ldots, e_{64}(\mathbf{x}_i,y)
\right]^{\mathsf T}
\in \mathbb{R}^{64}.
\end{equation}
Each component \(e_k(\mathbf{x}_i,y)\) is an abstract embedding coordinate, not a physical spectral band. Accordingly, \(\mathbf{e}(\mathbf{x}_i,y)\) should be interpreted as a learned latent summary of local surface context rather than as a single-pixel reflectance measurement.
In the downstream applications considered here, we use the embedding at an effective 1~km scale so that the extracted feature vector reflects surrounding landscape context rather than pixel-scale variation. This choice reduces sensitivity to exact centroid placement, local geolocation mismatch, and small-scale noise that is unlikely to be stably linked to sparse subsurface labels.

\par
This AlphaEarth representation provides the common surface feature space for both downstream applications studied in this paper. In the shallow \(V_{S}30\) application, it is evaluated both on its own and in combination with a small number of conventional terrain-based covariates, allowing comparison against standard predictors and hybrid feature sets. In the subsurface-temperature application, the same representation serves as the main learned surface input to a nonlinear model for thermal reconstruction. Taken together, these two applications examine how far a shared learned surface embedding can be carried into the subsurface across different physical settings, and whether a general-purpose geospatial representation can serve as an effective feature space for surface-to-subsurface inference.

\subsection{Shallow seismic velocities}
\par
The time-averaged shear-wave velocity in the upper 30 m, $V_S 30$, is the standard site descriptor for near-surface stiffness in earthquake engineering. 
It supports site classification, empirical site-amplification models, and rapid shaking products because it captures, in a single parameter, how shallow materials modify incoming seismic waves \citep{Borcherdt1994,GeyinMaurer2023}. 
Reliable $V_S 30$ maps are therefore needed for ground-motion modeling and seismic hazard analysis. 
At the same time, direct measurement remains costly. 
Boreholes, spectral-analysis-of-surface-waves (SASW), and other geophysical surveys provide high-quality point estimates, but the observations are sparse, method-dependent, and unevenly distributed at national scale \citep{McPhillipsEtAl2020}.

\par
Existing $V_S 30$ mapping approaches fall into three broad categories. 
The first is geology-based mapping, in which geologic units are assigned representative velocities and then refined where observations exist \citep{WillsEtAl2000}. 
The second is terrain-proxy mapping, especially the topographic-slope framework and its higher-resolution extensions, which exploit the empirical relation between relief, depositional environment, and shallow stiffness \citep{WaldAllen2007,AllenWald2009}. 
The third is hybrid or data-driven mapping, in which measured $V_S 30$ is combined with terrain, geology, or remote-sensing predictors using statistical or machine-learning models, primarily in research-and-development settings rather than routine operational deployment \citep{YongEtAl2012,ThompsonEtAl2014,GeyinMaurer2023}.

\par
Together, these studies show that land-surface properties carry useful information about shallow stiffness, but they also expose a persistent challenge. 
Slope, relief, geologic age, and other curated variables are informative, yet the relation between surface form and $V_S 30$ is scale-dependent and geographically variable \citep{LinEtAl2019}. 
A single topographic predictor can capture first-order mountain-basin contrasts, but it becomes less reliable where lithology, weathering, geomorphic history, vegetation, or human modification obscure that relation. 
\added[id=jl]{For example, the central Oregon coast is a representative location where slope alone can be misleading. 
Along this coastal belt, low-relief dunes, beach and estuarine sediments, fluvial deposits, marine terraces, and adjacent hillslopes occur in close proximity, so sites with similar slope can still have very different shallow material properties and therefore different $V_{S} 30$. 
The landscape also reflects complex depositional and erosional history, together with vegetation and local human modification, which further weakens the direct slope--$V_{S} 30$ relation. 
As a result, a slope-only model tends to smooth over strong local contrasts, whereas a richer geospatial representation is more useful.}

\subsubsection{AlphaEarth as a representation for \texorpdfstring{$V_S 30$}{VS 30} inference}
\par
In this study, we use AlphaEarth as a representation of landscape context to infer $V_S 30$. 
The physical connection is indirect but well grounded. 
Although $V_S 30$ itself is not observed from space, many of the processes that control shallow stiffness leave persistent surface expressions. 
Basin fill, regolith development, drainage organization, surface roughness, lithologic contrast, moisture regime, vegetation, and land use all influence optical, radar, topographic, and climate observations. 
AlphaEarth is well suited to this setting because it compresses this joint multi-source context into a 64-dimensional embedding field designed for transfer to downstream mapping tasks \citep{GoogleSatelliteEmbedding2025}. 
Rather than prescribing a fixed set of hand-crafted proxies, the embedding provides a compact description of coupled surface states that are likely to co-vary with shallow seismic velocity.

\par
In the implementation we use here, the 2024 annual AlphaEarth layer is sampled at each station location from the Google Earth Engine Satellite Embedding collection using a 1~km Earth Engine scale. 
We further supplement the embedding with topographic slope derived from USGS SRTM elevation \citep{FarrKobrick2000}, represented in log space, and with a binary west--east regional indicator, here termed tectonic status, defined by whether a site lies west or east of $105^\circ$W \added[id=jl]{(approximately west versus east of the Front Range of the Rocky Mountains). 
This tectonic-status division is intended as an empirical proxy for broad tectonic regime: regions west of the Rocky Mountain front and along the western continental margin are more commonly associated with tectonically younger and more active provinces, whereas areas east of the Rockies are generally part of the more stable continental interior.}
\added[id=jl]{This interpretation is also consistent with the USGS topographic-slope proxy framework, which identifies broad site-condition contrasts across the western United States, with faster materials associated with much of the Rocky and Cascade mountain ranges and slower materials interspersed within lower-relief basins \citep{WaldAllen2007}. 
Taken together, these features are included not to replace physical reasoning with a black-box representation, but to help the embedding capture landscape-scale and regional structure that simple descriptors alone cannot adequately represent.}

\subsubsection{Experimental setup}
\par
The target labels are taken from the \emph{Updated Compilation of $V_S 30$ Data for the United States} and are downloaded through the USGS $V_S 30$ portal \citep{McPhillipsEtAl2020}. 
This compilation assembles measured $V_S 30$ values and metadata from government-sponsored reports, online databases, and scientific and engineering journals, and includes both downhole and surface-based estimates. 
Although the portal includes sites outside the conterminous United States, we retain only conterminous U.S. stations in this study. 
After filtering and feature extraction, the final modeling dataset contains 2886 stations; these retained stations are the points shown in Figure~\ref{fig:vs30}a.
\added[id=jl]{Spatial heterogeneity appears as dense clusters in well-studied urban corridors and tectonically active western regions, with much sparser coverage in parts of the central and eastern United States. 
This pattern mainly reflects how data are collected: measurements are project-driven and cost-intensive, so surveys are preferentially conducted where seismic hazard, infrastructure exposure, or prior site investigations justify field campaigns, while remote or lower-priority areas remain under-sampled.}

\par
Downstream prediction is performed in $\log_{10}V_{S}30$ space using four feature sets: AlphaEarth alone, $\log_{10}$-slope plus tectonic status, AlphaEarth plus $\log_{10}$-slope, and AlphaEarth plus $\log_{10}$-slope plus tectonic status. 
We predict in $\log_{10}V_{S}30$ space with $\log_{10}$-slope because the relation between topographic slope and $V_{S}30$ is more naturally expressed on a logarithmic scale, consistent with previous studies \citep{WaldAllen2007}. 
Log scaling helps represent multiplicative variation, reduces compression of the low-slope range, and makes prediction errors more interpretable as relative rather than absolute differences. 
\par
We consider three regressors: ordinary linear regression, random forest \citep{Breiman2001}, and XGBoost \citep{ChenGuestrin2016}. 
All predictors are standardized before fitting. 
We use a random 80/20 train--test split, yielding 2308 training stations and 578 held-out test stations. 
Model selection is carried out only on the training subset using 3-fold cross-validation with shuffled folds. 
For linear regression, no hyperparameter tuning is performed. 
For random forest, we tune the number of trees, the maximum tree depth, and the minimum number of samples required at a leaf node. The candidate values are 200 or 400 trees, maximum depth of none, 25, or 50, and minimum leaf size of 1 or 5 samples. 
For XGBoost, we tune the number of boosting rounds, the learning rate, the maximum tree depth, the row subsampling fraction, the column subsampling fraction per tree, and the $L_2$ regularization strength. The candidate values are 200 or 400 boosting rounds, learning rate of 0.05 or 0.1, maximum depth of 4, 6, or 8, subsample fraction of 0.7 or 0.9, column subsample fraction of 0.6 or 0.8, and regularization strength of 1 or 5.
Hyperparameter tuning for random forest and XGBoost is performed by randomized search with up to 10 sampled parameter combinations per model--feature-set pair. 
To reduce sensitivity to stochastic variation in the ensemble-based methods, each random forest and XGBoost configuration is evaluated over five runs with different random seeds, and model selection is based on the average cross-validated performance across those runs. 
For each candidate configuration, the selection criterion is the mean cross-validated root mean squared error in $\log_{10}V_{S}30$ space. 
After identifying the best configuration, we refit that model on the full training subset and evaluate it on the held-out test set. 
Performance is reported in both physical and log space using root mean squared error (RMSE), mean absolute error (MAE), coefficient of determination ($R^2$), and coefficient of determination in log space ($R^2_{\log}$). 
The best-performing model is a random forest with 200 trees, a maximum tree depth of 25, a maximum feature fraction of 0.3 at each split.
After selecting the overall best-performing configuration, we retrain it on all 2886 available stations and apply it to a 1~km land grid across the conterminous United States.

\subsection{Subsurface temperature}
% \nori{Reduce the fraction of the training dataset. Also using only isolated states, not randomly from the entire US.}

Subsurface temperature distribution is critical for geothermal resource assessment, hydrocarbon exploration, and understanding crustal thermal states, yet borehole measurements remain sparse and unevenly distributed \citep{aljubran2024thermal}. Here we present a framework combining AlphaEarth foundation model embeddings with deep neural networks to reconstruct temperature distribution from 0 to 6000\,m depth across the conterminous United States. We first discretize the study area into a 20\,km regular grid and extract 64-dimensional embeddings for each grid cell using Google Earth Engine. A multilayer perceptron with batch normalization and dropout is then trained to establish nonlinear mapping between satellite embeddings and measured borehole temperatures from more than 400 thousand locations. The model achieves strong predictive performance ($R^2 = 0.919$, RMSE $= 6.0\,^\circ$C), substantially outperforming traditional machine learning method without embeddings. Feature importance analysis reveals that AlphaEarth embeddings capture integrated surface environmental signatures---including land surface temperature, vegetation phenology, topography, and geology---that correlate with deep geothermal gradients. Our approach demonstrates that AlphaEarth foundation models can effectively bridge surface observations and subsurface thermal regimes, providing a scalable pathway for high-resolution temperature mapping in data-sparse regions.
 
% \subsubsection{Overview}
 
Knowledge of subsurface temperature distribution is fundamental to understanding crustal thermal states, hydrocarbon exploration, and geothermal energy potential \citep{mather2022computer, benz2024global}. Temperature controls metamorphic reactions, fluid circulation, and rock rheology, influencing everything from earthquake behavior to ore deposit formation \citep{pollack1993heat}. For applied geosciences, accurate subsurface temperature maps guide geothermal exploration, enhance oil and gas production strategies, and inform carbon sequestration site selection \citep{blackwell2011temperature}.

Despite its importance, direct temperature measurements remain severely limited. Boreholes are expensive to drill, concentrated in energy-prospective regions, and rarely exceed two to three kilometers depth. The conterminous United States has hundreds of thousands of quality-controlled borehole temperature measurements, but these are unevenly distributed, being densely sampled in sedimentary basins and virtually absent in mountainous and remote regions. Traditional interpolation methods such as kriging produce smooth maps that fail to capture local thermal anomalies driven by complex geological controls.
 
Recent advances in satellite remote sensing and artificial intelligence offer new opportunities. The AlphaEarth Foundations model, developed by Google DeepMind, is a geospatial foundation model trained on billions of Earth observation data points including optical imagery, synthetic aperture radar, LiDAR, digital elevation models, and climate reanalysis \citep{BrownEtAl2025}. It generates 64-dimensional embeddings at ten-meter resolution that compress diverse surface information into analyzable representations. Unlike traditional spectral indices or engineered features, these embeddings are learned representations that capture both spatial context and temporal dynamics, including seasonal vegetation cycles, snow cover persistence, and surface thermal regimes.
 
The key insight for this section is that surface environmental signatures captured by AlphaEarth embeddings may correlate with deep geothermal gradients. Heat flow at depth is influenced by crustal radiogenic heat production, which is reflected in surface geology and geomorphology, and by hydrothermal circulation, which manifests in surface thermal anomalies and vegetation patterns. By learning these complex, nonlinear relationships, neural networks can potentially infer subsurface temperatures from surface observations alone.
 
In this section, we present a complete workflow involving grid-based discretization of the conterminous United States, extraction of AlphaEarth embeddings for each grid cell using Google Earth Engine, development of a neural network regressor trained on borehole temperature measurements, and generation of a continuous temperature map at one thousand meters depth. Our approach addresses the fundamental challenge of sparse subsurface data by leveraging globally available satellite embeddings, offering a scalable solution for temperature mapping in data-poor regions worldwide.
 
\subsubsection{Experimental setup}
\par
The temperature-mapping experiment is carried out over the conterminous United States ($24^\circ$N--$49.5^\circ$N, $125^\circ$W--$66^\circ$W). 
Target labels are compiled from the Southern Methodist University Geothermal Laboratory database, the USGS Geothermal Resources Assessment, and state geological survey records. 
The merged borehole archive contains more than 400{,}000 temperature measurements, but the spatial distribution is strongly uneven because drilling is concentrated in sedimentary basins, energy-producing provinces, and other regions of applied interest. 
Before modeling, we remove measurements affected by drilling disturbances, discard records with incomplete metadata, and retain only samples within the 0--6000\,m depth range. 
As in the $V_{S}30$ application, this sampling structure means that apparent skill must be interpreted with care because some of the prediction task still resembles interpolation within well-observed regions rather than strict extrapolation into data-poor terrain.

\par
To construct a national prediction domain, we discretize the conterminous United States into a regular 20\,km grid, yielding 20{,}971 grid cells. 
For each cell, we use the centroid coordinate for feature extraction and prediction. 
This spacing is chosen to balance computational cost against the need to resolve regional thermal gradients at continental scale. 
At each labeled borehole location and each grid centroid, we extract the 64-dimensional AlphaEarth embedding from the Google Earth Engine annual product and use that vector as the predictor set for downstream learning.

\par
We model temperature at 1000\,m depth using a multilayer perceptron that maps the 64-dimensional embedding to a scalar temperature prediction. 
The network consists of three hidden layers with 128, 64, and 32 neurons, respectively, each using ReLU activation; batch normalization is applied after the hidden layers, and 30\% dropout is used in the first two hidden layers to reduce overfitting. 
The output layer has a single neuron corresponding to predicted temperature. 
Model fitting minimizes mean squared error,
\begin{equation}
  \mathcal{L} = \frac{1}{N}\sum_{i=1}^{N}\!\left(T_i - \hat{T}_i\right)^{2},
\end{equation}
where $T_i$ is the measured temperature and $\hat{T}_i$ is the model prediction for sample $i$.

\par
The dataset is divided into training and test subsets using an 80/20 split with stratification by geographic region so that both subsets retain broad national coverage. 
Model selection is performed only on the training subset using 5-fold cross-validation, and final performance is reported on the held-out test set. 
We evaluate predictive skill using root mean squared error (RMSE), mean absolute error (MAE), and coefficient of determination ($R^2$). 
To interpret which embedding channels contribute most strongly to the temperature model, we also compute permutation importance \citep{Breiman2001} by randomly shuffling each embedding dimension across samples and measuring the corresponding increase in prediction error.
 
\section{Results}
\subsection{Near-surface seismic velocities}
\par
AlphaEarth accounts for most of the improvement over the terrain-only baseline, and the addition of simple geomorphic context yields the strongest overall performance. 
As summarized in Table~\ref{tab:vs30}, the model with the lowest test RMSE is the XGBoost model using AlphaEarth embeddings and $\log$-slope, achieving RMSE $=126.1$ m/s, MAE $=66.2$ m/s, $R^2=0.5329$, and $R^2_{\log}=0.6525$.
Relative to the conventional baseline using only $\log$-slope and tectonic status, this reduces test RMSE from 166.6 to 126.1 m/s, a 24.3\% improvement, and lowers MAE from 100.9 to 66.2 m/s, a 34.4\% improvement. 
At the same time, test $R^2$ increases from 0.1850 to 0.5329, nearly tripling the explained variance on the held-out set. 
AlphaEarth alone already performs substantially better than the terrain-only baseline, indicating that the embedding captures much of the site-condition information that conventional terrain proxies are intended to represent, while slope and tectonic status provide only a modest additional gain.

\par
To compare our method with an established community product, we also evaluate the USGS California $V_{S30}$ map~\citep{Thompson2018Vs30CA} on the California subset of our held-out test stations. 
On this subset, the best whole-US model (XGBoost with AlphaEarth embeddings and $\log$-slope) attains a similar RMSE to the sampled USGS map, while achieving lower MAE and substantially higher $R^2$ and $R^2_{\log}$. 
This comparison is not intended as a strict head-to-head assessment of predictive performance, because the USGS map is not a fully independent baseline: it was constructed by combining geologic and topographic constraints with site-specific $V_{S30}$ measurements through regression kriging, whereas our reported metrics are computed on unseen held-out stations. 
We therefore include this comparison primarily to situate our results relative to a widely used external benchmark and to show how the resulting California predictions compare quantitatively with an established community map.

\begin{table}[htbp]
\centering
\small
\begin{tabular}{llrrrrr}
\hline
Feature set & Model & $n$ features & Test RMSE & Test MAE & Test $R^2$ & Test $R^2_{\log}$ \\
\hline
AE+slope+tectonic status & XGBoost & 66 & 128.8 & 66.8 & 0.5131 & 0.6433 \\
AE+slope+tectonic status & RF & 66 & 127.4 & \textbf{66.0} & 0.5231 & 0.6513 \\
AE+slope+tectonic status & LinearReg & 66 & 150.1 & 87.9 & 0.3384 & 0.4410 \\
AE+slope & XGBoost & 65 & \textbf{126.1} & 66.2 & \textbf{0.5329} & \textbf{0.6525} \\
AE+slope & RF & 65 & 127.3 & 66.1 & 0.5239 & 0.6515 \\
AE+slope & LinearReg & 65 & 150.3 & 88.0 & 0.3367 & 0.4404 \\
AE & XGBoost & 64 & 131.5 & 67.5 & 0.4922 & 0.6309 \\
AE & RF & 64 & 131.0 & 67.0 & 0.4962 & 0.6328 \\
AE & LinearReg & 64 & 151.6 & 89.0 & 0.3248 & 0.4341 \\
slope+tectonic status & LinearReg & 2 & 166.6 & 100.9 & 0.1850 & 0.2692 \\
\hline
AE+slope (CA-only) & XGBoost & 65 & 130.0 & 73.2 & 0.5864 & 0.6887 \\
USGS map (CA-only) &  Baseline & -- & 129.8 & 87.8 & 0.3983 & 0.3623 \\
\hline
\end{tabular}
\caption{Comparison of model performance across feature sets and regression models for $V_{S30}$ prediction. Metrics are reported on the held-out test set for the full-data models. The final two rows report evaluation on the California-only subset of the same held-out test stations, comparing the best whole-US model against the established USGS California $V_{S30}$ map~\citep{Thompson2018Vs30CA}.}
\label{tab:vs30}
\end{table}

\par
\begin{figure}[htbp]
  \centering
  \includegraphics[width=\textwidth]{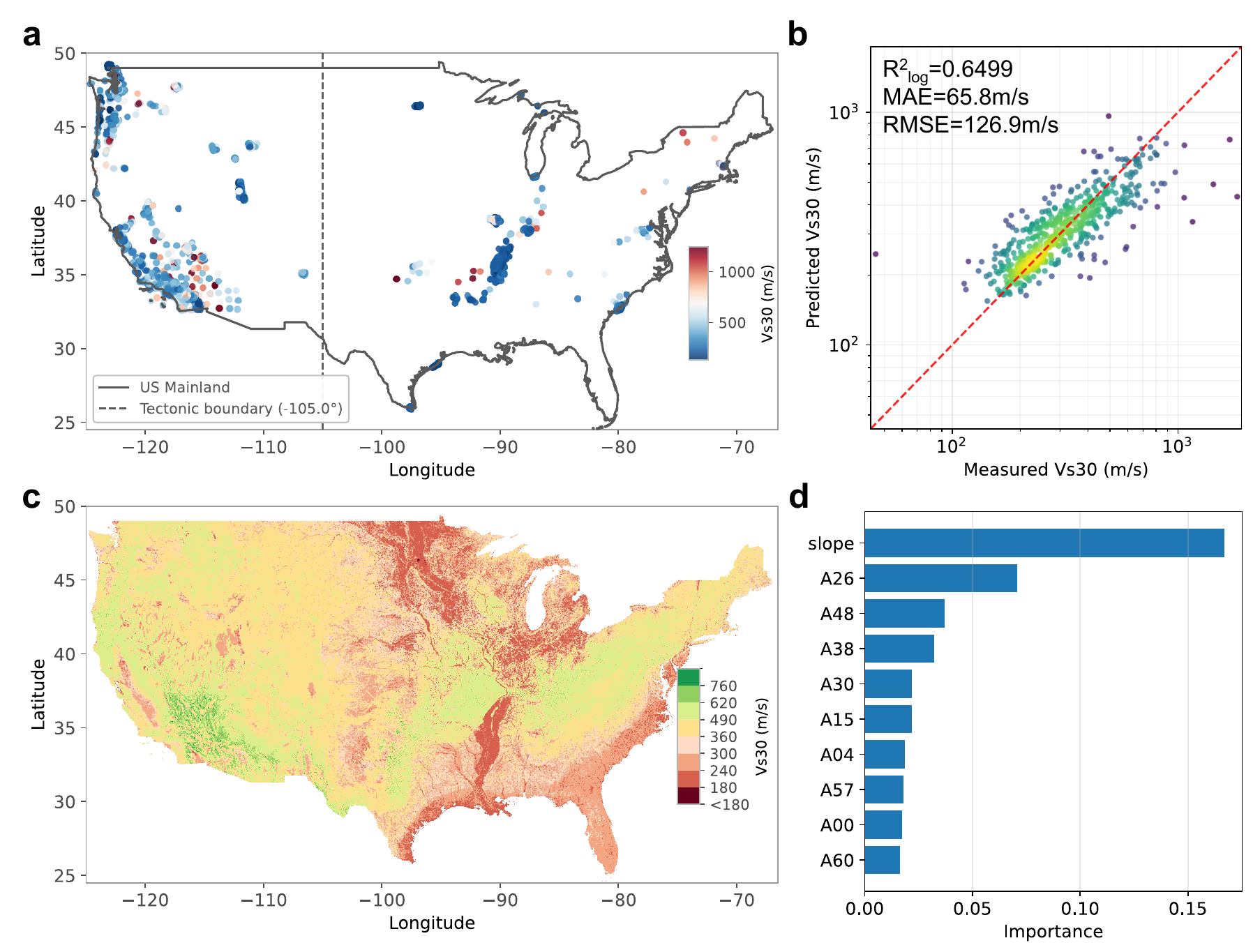}
  \caption{Overview of the $V_S 30$ dataset, model performance, national prediction, and model interpretation. 
  (a) Spatial distribution of the measured $V_S 30$ stations used in this study across the conterminous United States, colored by observed $V_S 30$, with the west-east tectonic boundary at $105^\circ$W. 
  (b) Comparison between predicted and measured $V_S 30$ for the held-out test set, shown on log-log axes with the 1:1 line. 
  (c) Predicted $V_S 30$ over a 1 km land grid across the conterminous United States using the best-performing random-forest model, shown as the resulting distribution of NEHRP site classes.
  (d) Top ten feature importances for the best-performing model, showing contributions from both $\log$-slope and multiple AlphaEarth embedding dimensions.}
  \label{fig:vs30}
\end{figure}

Figure~\ref{fig:vs30} provides a complementary view of the data, predictions, and model behavior. 
The upper-left panel shows the spatial distribution of the measured stations used in this study, colored by observed $V_S 30$, together with the west-east tectonic boundary at $105^\circ$W. 
Notably, the stations are concentrated in California and the western United States, with much sparser coverage across the central and eastern parts of the country. 
Figure~\ref{fig:vs30}b compares predicted and measured $V_S 30$ for the held-out test set on log-log axes. 
Most points cluster close to the 1:1 line, although the scatter increases toward the highest velocities. 
Figure~\ref{fig:vs30}c shows the 1 km gridded prediction over the conterminous United States, expressed as the distribution of National Earthquake Hazards Reduction Program (NEHRP) site classes based on the predicted $V_S 30$ field.
Lower predicted values are concentrated in the Gulf and Atlantic Coastal Plains, the Mississippi embayment, and the California Central Valley, whereas higher values are more common in the western uplands and other bedrock-dominated regions. 
Figure~\ref{fig:vs30}d ranks the ten most important features in the best random-forest model. 
$\log$-slope is the most important single variable, but several AlphaEarth dimensions also contribute strongly. 
This pattern suggests that the embedding is not simply acting as a surrogate for slope; rather, it provides additional information on landscape structure and surface condition that is relevant to shallow seismic stiffness.
\added[id=jl]{By contrast, tectonic status does not appear as an important feature in the importance ranking. 
Because tectonic status is defined only as a binary west--east regional split, it provides a very coarse description of tectonic setting, whereas both the AlphaEarth embedding and $\log$-slope contain substantially richer spatial information. 
As a result, whatever first-order regional signal tectonic status carries may already be represented by these other variables, leaving it with relatively little additional predictive value in the model.}

\subsection{Subsurface temperature}
 
\begin{figure}[htbp]
  \centering
  \includegraphics[width=\textwidth]{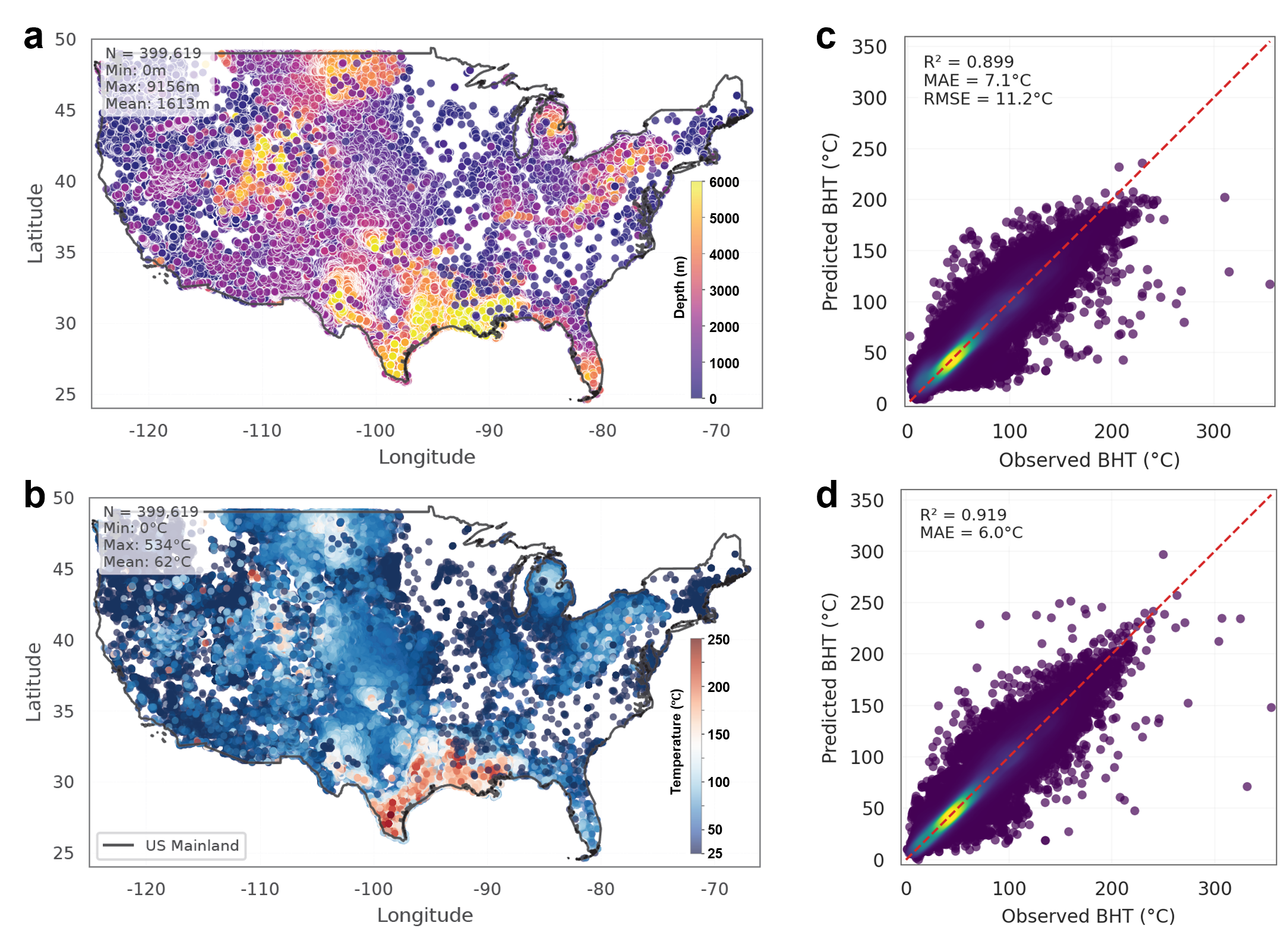}
  \caption{\textbf{(a)} Depth spatial distribution of boreholes from
    the database. \textbf{(b)} Observed bottom-hole temperature
    distribution of boreholes from the database. \textbf{(c)} Observed
    versus predicted bottom-hole temperature distribution for the
    machine learning method \emph{without} AlphaEarth embeddings.
    \textbf{(d)} Observed versus predicted bottom-hole temperature
    distribution for the machine learning method \emph{with} AlphaEarth
    embeddings.}
  \label{fig:borehole}
\end{figure}

AlphaEarth embeddings are successfully extracted for 20,971 grid cells covering the conterminous United States. Extraction succeeded for 98.7\% of cells; failures occurred primarily in coastal boundary cells where centroid coordinates fell outside the embedding image domain. Figure \ref{fig:borehole}a--b shows the distribution of extracted points and borehole locations for depth and temperature distributions.
 
Figure \ref{fig:borehole}c--d illustrates the observed bottom-hole temperature distribution versus predicted bottom-hole temperature distribution for the machine learning method with and without AlphaEarth embeddings. Results reveal that the first three components explain 47\% of total variance, suggesting that embedding dimensions capture complementary information rather than being highly redundant.
 
\begin{figure}[htbp]
  \centering
  \includegraphics[width=\textwidth]{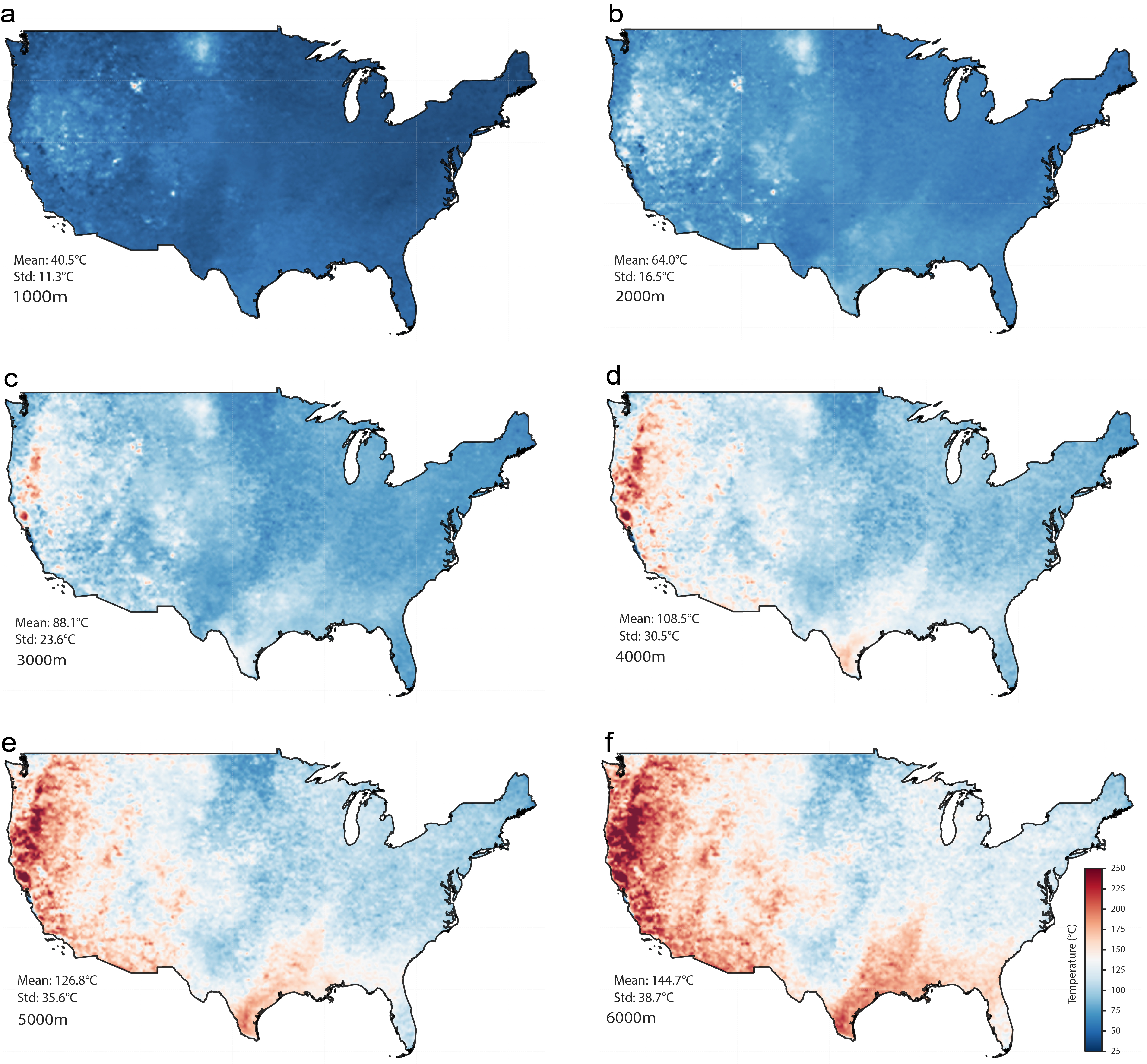}
  \caption{Subsurface temperature reconstruction using AlphaEarth
    foundation model enabled neural network with depth ranging from
    1000\,m to 6000\,m.
    }
  \label{fig:tempmap}
\end{figure}

Applying the trained neural network to all grid cells generated a continuous temperature map with depth ranging from 1000\,m to 6000\,m across the conterminous United States (Figure~\ref{fig:tempmap}). The reconstructed field shows a broad first-order contrast between the tectonically active western United States and the more stable continental interior. The highest predicted temperatures occur in the western United States, especially across the Basin and Range province, along the Yellowstone hotspot track, and in other volcanically and tectonically active regions, where elevated heat flow and crustal deformation are consistent with warmer subsurface conditions. Intermediate temperatures are more common in major sedimentary basins and along tectonically active margins, where subsurface thermal structure reflects a combination of burial history, fluid circulation, and regional tectonic setting. In contrast, the lowest predicted temperatures are concentrated in the stable cratonic regions of the central and eastern United States, particularly across parts of the Upper Midwest and New England, where old lithosphere, lower heat flow, and reduced tectonic activity are broadly consistent with cooler thermal regimes at depth.  The map resolves thermal anomalies at scales much finer than possible with interpolation alone, such as temperature gradients across fault-bounded basins and geothermal signatures associated with Quaternary volcanism.
 
\section{Discussion}
\par
A central contribution of this study is to evaluate the Alphaearth foundation-model surface representation across two distinct subsurface inference problems: shallow mechanical structure represented by $V_{S}30$ and deeper thermodynamic structure represented by subsurface temperature. In the discussion below, we first consider the implications and limitations of each application separately, and then synthesize their similarities and differences to identify broader methodological lessons for surface-to-subsurface mapping.
\subsection{Near-surface seismic velocities}
\par
The $V_S 30$ estimation with AlphaEarth embedding has practical implications for regional seismic hazard analysis. 
Traditional $V_S 30$ mapping often relies on slope because it is globally available and physically interpretable, but its relation to shallow stiffness weakens where lithology, weathering history, sediment thickness, vegetation, or urban modification disturb simple topographic controls.
The present results in this study indicate that AlphaEarth can recover part of this missing context while still benefiting from an explicit slope term. 
In practice, it offers a scalable prior for regional mapping and screening applications, especially in areas where direct measurements are sparse and detailed geologic inputs are unavailable or inconsistent. 
It does not replace site-specific investigations for engineering design, but it can improve the quality and spatial scalabiltiy of $V_S 30$ products. 

\par
Several limitations should nonetheless be kept in mind. 
The USGS compilation merges measurements from many studies, so part of the residual variance likely reflects differences in acquisition method, processing workflow, and spatial support rather than only deficiencies in the predictors. 
The station distribution is also highly uneven, with dense clusters in California and the western United States and much sparser coverage across the central and eastern states. 
Because the present evaluation uses a random train-test split, some of the reported skill likely reflects interpolation within the existing observation footprint rather than strict geographic extrapolation. 

\par
These limitations point to several natural next steps. 
A more stringent assessment of generalization would use spatial block cross-validation, leave-one-state-out tests, or evaluation by physiographic province. 
Multi-scale pooling of AlphaEarth features around each station may better separate local site response from broader geomorphic setting than single-point sampling. 
The framework could also be extended with explicit geology, roughness, curvature, and hydrographic descriptors, and coupled with uncertainty-aware learners so that the output is a probabilistic $V_S 30$ surface rather than a single deterministic estimate. 
A further avenue is interpretability: identifying what the most informative embedding dimensions represent in physical terms would help connect the predictive gains back to process-based understanding. 
More broadly, this case study suggests that foundation-model embeddings may provide a useful bridge between Earth observation and engineering seismology, not only for $V_S 30$, but for other near-surface properties that are difficult to observe directly yet leave persistent signatures in the landscape.

\subsection{Subsurface temperatures}
\par
This work demonstrates that satellite foundation models can bridge surface observations and subsurface properties, at greater depth than in the near-surface seismic-velocity application. A plausible physical explanation is that tectonically active regions often exhibit rougher topography, higher relief, and stronger landscape disequilibrium, and these same regions commonly coincide with elevated heat flow and higher subsurface temperatures  \citep{marzotto2025olivine}. In contrast, tectonically stable cratonic regions are often topographically smoother and thermally cooler. Although this linkage is indirect and modulated by lithology, sedimentation, and hydrology, AlphaEarth embeddings can encode these coupled surface expressions and thus provide useful constraints for thermal inference. Beyond temperature mapping itself, the approach has broader implications for geothermal exploration through rapid and lower-cost regional screening before intensive field campaigns, and for carbon sequestration by helping identify areas with thermal regimes that are more compatible with subsurface CO$_2$ storage objectives.

The same framework may also be adaptable to planetary applications, where orbital surface observations could be used to infer shallow subsurface conditions on bodies such as Mars or Venus. Future work should integrate additional geophysical constraints (e.g., gravity, magnetics, and seismicity) with AlphaEarth embeddings, extend predictions to multiple depths using physics-informed models, and validate results with new borehole measurements in under-sampled regions.

\subsection{Similarities and differences}
The two applications follow a shared methodological pattern: both use AlphaEarth embeddings as transferable geospatial representations and combine them with supervised learning to infer subsurface properties, namely elastic stiffness and temperature, that are not directly measured from satellite imagery. In both cases, adding embedding information improves predictive skill over simpler baselines, indicating that coupled surface signals carry meaningful constraints on subsurface structure and state. Both applications also depend on sparse and heterogeneous in situ labels, and both upscale site-level learning to continuous regional maps.

The key differences arise from the target properties and the corresponding modeling choices. The $V_S 30$ application focuses on a shallow mechanical descriptor of near-surface stiffness, where physically interpretable proxies such as slope and regional tectonic partitioning remain highly informative. The subsurface-temperature application targets subsurface temperature, a deeper thermodynamic field with broader spatial coherence and more diffuse controls, and therefore uses a neural-network mapping to capture nonlinear interactions among embedding dimensions. As a result, the $V_S 30$ application emphasizes hybrid feature design and incremental gains over terrain-only baselines, while the temperature application emphasizes large-scale reconstruction capability and spatial thermal-pattern recovery.

\par
The two sections also differ in interpretability strategy. For $V_S 30$, the feature-importance ranking supports a complementary view: classical geomorphic predictors remain central, but several embedding dimensions add nonredundant predictive content. For temperature, interpretation is more indirect because latent nonlinear representations are less transparent; however, the performance gain suggests that the embedding encodes integrated surface context relevant to deep thermal regimes. Taken together, the comparison suggests a practical principle: where strong physics-guided proxies exist, embeddings are most effective as augmenting features; where controls are multi-factor and weakly parameterized, embeddings can serve as the primary representation.

A common limitation across both applications is potential overestimation of generalization due to sampling structure. Uneven station distribution, mixed measurement protocols, and random train-test splits can inflate apparent skill through spatial interpolation within dense observation regions. This concern affects both tasks, although it may manifest differently: for $V_S 30$ as regional bias in site-class boundaries, and for temperature as oversmoothed gradients in under-sampled provinces. Future work should apply a unified validation framework across both sections, including spatial block cross-validation, leave-region-out experiments, and explicit uncertainty calibration.

Overall, the two applications are scientifically distinct but methodologically consistent. The $V_S 30$ application shows that AlphaEarth improves near-surface seismic site characterization when fused with established domain predictors, whereas the temperature application shows that the same representation supports continental-scale thermal reconstruction at depth. Their agreement across different physics domains strengthens the central claim of this study: foundation-model embeddings are most valuable not as replacements for geoscientific reasoning, but as scalable shared representations that can be adapted to multiple subsurface targets when paired with task-appropriate models and rigorous validation.

\section{Conclusions}
\par
This study evaluates whether AlphaEarth embeddings, trained by surface data, can support scalable inference of subsurface properties through two applications: shallow seismic velocity and subsurface temperature mapping. Across both tasks, embedding-informed models outperform simpler baselines, indicating that multi-source surface context provides useful constraints on subsurface variability when paired with task-appropriate learning frameworks. We also find that non-embedding covariates remain valuable for stabilizing regression estimates, for example log-slope and tectonic-status terms in the $V_S 30$ model. Despite uneven label coverage and heterogeneous measurements, the models recover meaningful and physically consistent regional patterns, while potential spatial leakage under random data splits remains an important caveat for interpretation. Future work should prioritize spatially robust validation, uncertainty quantification, and integration with physics-guided constraints for reliable prediction.

\section*{Acknowledgments}
We thank the Google DeepMind for making the AlphaEarth embeddings publicly available, which enabled this study.

%Bibliography
\bibliographystyle{abbrvnat}
\bibliography{references}

@book{mather2022computer,
  title={Computer processing of remotely-sensed images},
  author={Mather, Paul M and Koch, Magaly},
  year={2022},
  publisher={John Wiley \& Sons}
}

@article{marzotto2025olivine,
  title={Olivine’s high radiative conductivity increases slab temperature by up to 200K},
  author={Marzotto, Enrico and Koptev, Alexander and Speziale, Sergio and Koch-M{\"u}ller, Monika and Abdel-Hak, Nada and Cichy, Sarah B and Lobanov, Sergey S},
  journal={Nature Communications},
  volume={16},
  number={1},
  pages={6058},
  year={2025},
  publisher={Nature Publishing Group UK London}
}

@article{aljubran2024thermal,
  title={Thermal Earth model for the conterminous United States using an interpolative physics-informed graph neural network},
  author={Aljubran, Mohammad J and Horne, Roland N},
  journal={Geothermal Energy},
  volume={12},
  number={1},
  pages={25},
  year={2024},
  publisher={Springer}
}

@article{Borcherdt1994,
  author = {Borcherdt, Roger D.},
  title = {Estimates of Site-Dependent Response Spectra for Design (Methodology and Justification)},
  journal = {Earthquake Spectra},
  volume = {10},
  number = {4},
  pages = {617-653},
  year = {1994},
  doi = {10.1193/1.1585791}
}

@article{WillsEtAl2000,
  author = {Wills, Chris J. and Petersen, Mark D. and Bryant, William A. and Reichle, Michael S. and Saucedo, George J. and Tan, Siang and Taylor, George C. and Treiman, Jerome A.},
  title = {A Site-Conditions Map for California Based on Geology and Shear-Wave Velocity},
  journal = {Bulletin of the Seismological Society of America},
  volume = {90},
  number = {6B},
  pages = {S187-S208},
  year = {2000},
  doi = {10.1785/0120000503}
}

@article{WaldAllen2007,
  author = {Wald, David J. and Allen, Trevor I.},
  title = {Topographic Slope as a Proxy for Seismic Site Conditions and Amplification},
  journal = {Bulletin of the Seismological Society of America},
  volume = {97},
  number = {5},
  pages = {1379-1395},
  year = {2007},
  doi = {10.1785/0120060267}
}

@article{AllenWald2009,
  author = {Allen, Trevor I. and Wald, David J.},
  title = {On the Use of High-Resolution Topographic Data as a Proxy for Seismic Site Conditions (VS30)},
  journal = {Bulletin of the Seismological Society of America},
  volume = {99},
  number = {2A},
  pages = {935-943},
  year = {2009},
  doi = {10.1785/0120080255}
}

@article{YongEtAl2012,
  author = {Yong, Alan and Hough, Susan E. and Iwahashi, Junko and Braverman, Amy},
  title = {A Terrain-Based Site-Conditions Map of California with Implications for the Contiguous United States},
  journal = {Bulletin of the Seismological Society of America},
  volume = {102},
  number = {1},
  pages = {114-128},
  year = {2012},
  doi = {10.1785/0120100262}
}

@article{ThompsonEtAl2014,
  author = {Thompson, Eric M. and Wald, David J. and Worden, C. Bruce},
  title = {A VS30 Map for California with Geologic and Topographic Constraints},
  journal = {Bulletin of the Seismological Society of America},
  volume = {104},
  number = {5},
  pages = {2313-2321},
  year = {2014},
  doi = {10.1785/0120130312}
}

@article{LinEtAl2019,
  author = {Lin, Jessica C. and Moon, Seulgi and Yong, Alan and Meng, Lingsen and Davis, Paul M.},
  title = {Length-Scale-Dependent Relationships between VS30 and Topographic Slopes in Southern California},
  journal = {Bulletin of the Seismological Society of America},
  volume = {109},
  number = {6},
  pages = {2614-2625},
  year = {2019},
  doi = {10.1785/0120190076}
}

@article{GeyinMaurer2023,
  author = {Geyin, Mertcan and Maurer, Brett W.},
  title = {U.S. National VS30 Models and Maps Informed by Remote Sensing and Machine Learning},
  journal = {Seismological Research Letters},
  volume = {94},
  number = {3},
  pages = {1467-1477},
  year = {2023},
  doi = {10.1785/0220220181}
}

@article{BrownEtAl2025,
  author = {Brown, Christopher F. and Kazmierski, Michal R. and Pasquarella, Valerie J. and others},
  title = {AlphaEarth Foundations: An Embedding Field Model for Accurate and Efficient Global Mapping from Sparse Label Data},
  journal = {arXiv preprint arXiv:2507.22291},
  year = {2025},
  doi = {10.48550/arXiv.2507.22291}
}

@misc{GoogleSatelliteEmbedding2025,
  author = {{Google Earth Engine and Google DeepMind}},
  title = {Satellite Embedding V1},
  year = {2025},
  howpublished = {Google Earth Engine Data Catalog},
  url = {https://developers.google.com/earth-engine/datasets/catalog/GOOGLE_SATELLITE_EMBEDDING_V1_ANNUAL},
  note = {Accessed 2026-03-15}
}

@article{FarrKobrick2000,
  author = {Farr, Tom G. and Kobrick, Michael},
  title = {Shuttle Radar Topography Mission Produces a Wealth of Data},
  journal = {Eos, Transactions American Geophysical Union},
  volume = {81},
  number = {48},
  pages = {583-585},
  year = {2000}
}

@article{Breiman2001,
  author = {Breiman, Leo},
  title = {Random Forests},
  journal = {Machine Learning},
  volume = {45},
  number = {1},
  pages = {5-32},
  year = {2001},
  doi = {10.1023/A:1010933404324}
}

@inproceedings{ChenGuestrin2016,
  author = {Chen, Tianqi and Guestrin, Carlos},
  title = {XGBoost: A Scalable Tree Boosting System},
  booktitle = {Proceedings of the 22nd ACM SIGKDD International Conference on Knowledge Discovery and Data Mining},
  pages = {785-794},
  year = {2016},
  doi = {10.1145/2939672.2939785}
}

@misc{McPhillipsEtAl2020,
  author = {McPhillips, Devin F. and Herrick, Julie A. and Ahdi, Sean K. and Yong, Alan K. and Haefner, Scott},
  title = {Updated Compilation of VS30 Data for the United States},
  year = {2020},
  publisher = {U.S. Geological Survey},
  doi = {10.5066/P9H5QEAC},
  note = {U.S. Geological Survey data release}
}

@article{benz2024global,
  title={Global groundwater warming due to climate change},
  author={Benz, Susanne A and Irvine, Dylan J and Rau, Gabriel C and Bayer, Peter and Menberg, Kathrin and Blum, Philipp and Jamieson, Rob C and Griebler, Christian and Kurylyk, Barret L},
  journal={Nature Geoscience},
  volume={17},
  number={6},
  pages={545--551},
  year={2024},
  publisher={Nature Publishing Group UK London}
}

@techreport{blackwell2011temperature,
  title={Temperature-at-depth maps for the conterminous US and geothermal resource estimates},
  author={Blackwell, David and Richards, Maria and Frone, Zachary and Batir, Joe and Ruzo, Andr{\'e}s and Dingwall, Ryan and Williams, Mitchell},
  year={2011},
  institution={Southern methodist university geothermal laboratory, Dallas, TX (United States)}
}

@article{pollack1993heat,
  title={Heat flow from the Earth's interior: analysis of the global data set},
  author={Pollack, Henry N and Hurter, Suzanne J and Johnson, Jeffrey R},
  journal={Reviews of Geophysics},
  volume={31},
  number={3},
  pages={267--280},
  year={1993},
  publisher={Wiley Online Library}
}

@inproceedings{fang2026yield,
  author    = {Fang, Jichao and Wu, Mingda and Zhang, Zhou and Luo, Wei},
  title     = {Application of AlphaEarth GeoFoundation Model for High-Accuracy Crop Yield Estimation},
  booktitle = {Proceedings of the IEEE/CVF Winter Conference on Applications of Computer Vision (WACV) Workshops},
  year      = {2026},
  month     = {March},
  pages     = {1465--1474}
}

@article{ryan2026wetland,
  title   = {Streamlining Wetland Vegetation Mapping with AlphaEarth Embeddings: Comparable Accuracy to Traditional Methods with Cleaner Maps and Minimal Preprocessing},
  author  = {Ryan, Shawn and Powell, Megan and Ling, Joanne and Wen, Li},
  journal = {Remote Sensing},
  year    = {2026},
  volume  = {18},
  number  = {2},
  pages   = {293},
  doi     = {10.3390/rs18020293}
}

@article{jin2026biomass,
  title   = {Assessing the Utility of Satellite Embedding Features for Biomass Prediction in Subtropical Forests with Machine Learning},
  author  = {Jin, Chao and Jiang, Xiaodong and Wen, Lina and Wu, Chuping and Xu, Xia and Jiao, Jiejie},
  journal = {Remote Sensing},
  year    = {2026},
  volume  = {18},
  number  = {3},
  pages   = {436},
  doi     = {10.3390/rs18030436}
}

@article{pascual2026biomass,
  title   = {Integration of Google's Alpha Earth Foundations into biomass estimation combined with GEDI spaceborne lidar and field inventory data},
  author  = {Pascual, Adrian and Guerra-Hernandez, Juan},
  journal = {Forest Ecology and Management},
  year    = {2026},
  volume  = {606},
  pages   = {123550},
  doi     = {10.1016/j.foreco.2026.123550}
}

@article{alvarez2025air,
  title   = {Machine Learning for Urban Air Quality Prediction Using Google AlphaEarth Foundations Satellite Embeddings: A Case Study of Quito, Ecuador},
  author  = {Alvarez, Cesar Ivan and Ulloa Vaca, Carlos Andres and Echeverria Llumipanta, Neptali Armando},
  journal = {Remote Sensing},
  year    = {2025},
  volume  = {17},
  number  = {20},
  pages   = {3472},
  doi     = {10.3390/rs17203472}
}

@misc{Thompson2018Vs30CA,
  author       = {Thompson, E. M.},
  title        = {An Updated {V}s30 Map for California with Geologic and Topographic Constraints},
  year         = {2018},
  note         = {U.S. Geological Survey data release, ver. 2.0, July 2022},
  doi          = {10.5066/F7JQ108S},
  url          = {https://doi.org/10.5066/F7JQ108S}
}

\end{document}